\documentclass[12pt]{article}
\usepackage{multirow}
\usepackage{jheppub}
\usepackage{graphicx,psfrag, subfigure}
\usepackage{amsmath}
\usepackage{dsfont,caption}
\usepackage{setspace}
\usepackage{soul}
\numberwithin{equation}{section}
\usepackage{relsize}
\usepackage{hyperref}
\usepackage{amssymb}
\newcommand{\bs}{\boldsymbol}

\DeclareMathAlphabet\mathbfcal{OMS}{cmsy}{b}{n}
\DeclareSymbolFont{bbold}{U}{bbold}{m}{n}
\DeclareSymbolFontAlphabet{\mathbbold}{bbold}
\DeclareSymbolFontAlphabet{\mathbbold}{bbold}

\makeatletter
\newcommand{\fixed@sra}{$\vrule height 2\fontdimen22\textfont2 width 0pt\shortrightarrow$}
\newcommand{\shortarrow}[1]{%
  \mathrel{\text{\rotatebox[origin=c]{\numexpr#1*45}{\fixed@sra}}}
}

\def\simleq{\; \raise0.3ex\hbox{$<$\kern-0.75em
      \raise-1.1ex\hbox{$\sim$}}\; }
\def\simgeq{\; \raise0.3ex\hbox{$>$\kern-0.75em
      \raise-1.1ex\hbox{$\sim$}}\; }
  \newcommand{\secref}[1]{Section \ref{#1}}

\def\M{M_{{\rm{Pl}}}}

\def\be{\begin{equation}}
\def\ee{\end{equation}}
\def\bea{\begin{eqnarray}}
\def\eea{\end{eqnarray}}

\def\be{\begin{equation}}
\def\ee{\end{equation}}
\def\bea{\begin{eqnarray}}
\def\eea{\end{eqnarray}}

\definecolor{mkcolor}{rgb}{1, .83,.83}
\definecolor{tbcolor}{rgb}{.83, .9,.83}
\definecolor{ojcolor}{rgb}{.83,.83, 1}
\definecolor{kecolor}{rgb}{1, 1, .5}

\usepackage{marginnote}

\begin{document}
\begin{titlepage}
\setcounter{page}{1} \baselineskip=15.5pt \thispagestyle{empty}

\bigskip\
\begin{center}

{\Large \bf The Axidental Universe}
\vskip 5pt

\vskip 15pt
\end{center}
\vspace{0.5cm}
\begin{center}
 
{Thomas C. Bachlechner$^*$, Kate Eckerle$^{ \dagger, \ddagger}$, Oliver Janssen$^\natural$ and Matthew Kleban$^\natural$}
\vspace{0.5cm}

\textsl{$^*$Department of Physics, University of California San Diego, La Jolla, USA} \\ \vspace{0.2cm}
\textsl{$ ^\dagger$Dipartimento di Fisica, Universit\`a di Milano-Bicocca, Milan, Italy}\\ \vspace{0.2cm}
\textsl{$^\ddagger$INFN, sezione di Milano-Bicocca, Milan, Italy}\\ \vspace{0.2cm} 
\textsl{$^\natural$Center for Cosmology and Particle Physics, New York University, New York, USA}
\end{center}
\vspace{-0.5cm}
{\small \noindent \\
\begin{center}
	\textbf{Abstract}

\end{center}
	Theories with several hundred axion fields have enormous numbers of distinct meta-stable minima.  A small fraction of these local minima have vacuum energy compatible with current measurements of dark energy.  The potential also contains regions suitable for inflation, and gives rise to a natural type of dark matter.  First-order phase transitions from one minimum to the vicinity of another play the role of big bangs and produce many bubbles containing evolving Friedmann-Lema\^itre-Robertson-Walker universes. The great majority either  collapse in a tiny fraction of a second, or expand exponentially forever as empty, structureless universes.  However, restricting to those bubble universes that form non-linear structure at some time in their history we find cosmologies that look remarkably similar to ours.  They undergo about 60 efolds of inflation, making them flat, homogeneous and isotropic, and endowing them with a nearly scale-invariant spectrum of primordial  density perturbations with roughly the observed magnitude and tilt.  They reheat after inflation to a period of radiation domination, followed by matter domination with roughly the observed abundance, followed by vacuum energy domination at roughly the observed density.  None of these features require any model building or  small parameters.  Instead, all dimensionful parameters in the theory can be set equal to the grand unified scale $10^{-2} \M$, and the dimensionless parameters are order one and can be chosen randomly.  The   small value of dark energy ultimately comes from non-perturbative gravitational effects, giving $\rho_{\text{DE}} \approx \Lambda^4 \, e^{-\mathcal{O}(1)\times \M/f}$, where $f \approx \Lambda \approx 10^{-2} \M$ are parameters of the axion theory.  Therefore, random axion landscapes can account for many of the apparently tuned features of our universe, including its current enormous size, age, and tiny energy densities compared to the scales of fundamental physics.
\noindent}

\vfill

\pagenumbering{gobble}
\begin{flushleft}
\vspace{0.8cm} \small \today
\end{flushleft}
\end{titlepage}
\tableofcontents

\newpage
\pagenumbering{arabic}
\section{Introduction}\label{intro}
Arguably the biggest problem in fundamental physics is the cosmological constant (CC) problem: why  the observed dark energy density $\rho_{\text{DE}}$ is so much smaller than the very large value naturally predicted by quantum field theory.   Since dark energy dominates the universe today and therefore determines the radius of the cosmological horizon, the CC problem corresponds to the question of why today's universe is so large compared to the length scales of fundamental physics, a question with a long history \cite{DiracCC,davies1982accidental}.

A promising solution to the CC problem is the existence of a large potential energy landscape with ${\cal N}_{\text{vac}}$ distinct meta-stable minima (also referred to as ``vacua'' or ``phases"), each with its own vacuum energy density, so that there is a discrete but very finely-spaced set of possible vacuum energies.  If the  vacuum energy densities are roughly uniformly distributed from $-\Lambda^4$ to $+\Lambda^4$ for some high energy $\Lambda$,\footnote{We use natural units, where energy density has units of energy$^4$.} the existence of \emph{some} vacua with vacuum energies consistent with observation requires that the number of minima ${\cal N}_{\text{vac}}$ be very large: 
\be \label{Nv}
{\cal N}_{\text{vac}} \simgeq \Lambda^4/\rho_{\text{DE}} \approx 10^{120} \,,
\ee
where the $\approx$ holds if $\Lambda \approx \M$.

Very large  ${\cal N}_{\text{vac}}$ can arise in theories where the vacuum is determined by multiple independent discrete choices (for instance, the choice of $\mathcal{O}(100)$ independent, bounded integers \cite{Bousso:2000xa}).  Such choices
 seem to be a feature of string theory, where a typical compactification to four large dimensions comes with hundreds of moduli fields and flavors of flux.   As we will see, large ${\cal N}_{\text{vac}}$ also arises in the four-dimensional effective theories of multiple axion fields  that capture some of the physics of such compactifications.  In fact, the existence of such huge numbers of meta-stable configurations seems to be a generic feature of the low-energy world (two examples are spin glasses and protein folding).  The basic assumption needed here is that the same type of phenomenon occurs at much higher energy scales.

The vast majority of these minima have energy at the natural scale of the theory, $\rho_{\text{vacuum}} \approx \Lambda^4 \gg \rho_{\text{DE}}$.  The explanation for why the region around us is in a special phase  of very low vacuum energy density is that the high energy in a typical phase causes either rapidly accelerated expansion that prohibits any dense structures from forming (if $\rho_{\text{vacuum}} > 0$), or collapse into a singularity in time too short for life to evolve (if $\rho_{\text{vacuum}} < 0$) \cite{Weinberg:1987dv,Tegmark:2005dy}. Only those minima with vacuum energies in the narrow range $|\rho_\text{vacuum}| \lesssim \text{few} \times \rho_\text{DE}  $ can contain observers. It is widely believed that all minima -- including those in this range, assuming there are any -- will eventually be populated by first-order phase transitions and hence exist somewhere in the universe, regardless of the initial conditions \cite{Brown:2011ry}.

However,  the condition \eqref{Nv} is not \emph{sufficient} for structure formation.   For instance, suppose a landscape satisfies ${\cal N}_{\text{vac}} \gg 10^{120}$, but does not allow for slow roll inflation.  If the  minima in this landscape are populated by Coleman-de Luccia tunneling from other (higher energy) minima, the negative curvature inside the bubbles will prohibit structure formation regardless of the value of dark energy \cite{Vilenkin:1996ar,Garriga:1998px,Freivogel:2005vv}.   This prevents the toy landscape of \cite{ArkaniHamed:2005yv} from explaining the observed value of the CC, even if it  satisfies   \eqref{Nv}.

In fact, \eqref{Nv}  is also not \emph{necessary} for structure formation.  Suppose the universe became matter dominated at a much larger energy density than occurred in our universe.  This would occur if the dark matter density were not for some reason much smaller than the radiation density in the early universe.  Since perturbations grow  rapidly during matter domination, such universes form collapsed structures much earlier and at much higher energy densities than in our universe,  long before the observed dark energy density  $\rho_{\text{DE}}$ would have any effect. Observers in such a universe would therefore expect to observe much larger values of the vacuum energy than we do, and such larger values could be accommodated by a smaller value of ${\cal N}_{\text{vac}}$. 

\emph{Therefore,  ${\cal N}_{\text{vac}} \gg 10^{120}$ is  neither necessary nor sufficient  to provide an explanation for the large-scale characteristics of our universe.} Instead, one must consider the set of cosmological histories in the  theory, discard  those that do not form structure, and  ask what the typical values of $\rho_{\text{DE}}$ and other observables are within the set that do form structure.  Due to the enormous complexity of any landscape with ${\cal N}_{\text{vac}} \gg 10^{120}$, this sounds like a daunting task.  However, the techniques we have developed in previous work \cite{bejk1,bejk2,bejk3} will make it possible in a large class of multi-axion theories.

Throughout this work we will take the approach of analyzing as large and complete a set of cosmological histories as we are able to with our techniques.  Within that set we will isolate those that form structure (defined by matter perturbations reaching $\mathcal{O}(1)$ at some time in the history) and describe their characteristics.  A more complete analysis would include effects from the ``measure'' that (for instance) might take the differing decay rates of the parent vacua into account (see \cite{Garriga:2005av,Bousso:2006ev,Hartle:2016tpo} for several differing approaches to the measure, or \cite{Freivogel:2011eg} and the references therein for a review, and Section 2 and the appendices of \cite{bejk3} for an analysis of decay rates in multi-axion theories).  We leave such an analysis to the future.

\subsection{Axion landscapes}
In this paper we describe the set of cosmological histories in theories of multiple axions. Axions are pseudo-scalar fields that possess a continuous shift symmetry (constant potential energy) that is exact under all perturbative quantum corrections. Hence, flat directions are only broken by non-perturbative effects. These adhere to a very specific form: they break the otherwise independent continuous shift symmetries to a set of discrete ones. In other words, the potential energy is a periodic function of the axion fields.

Axion fields are a common by-product of string compactifications. When string theory is compactified to four dimensions, the resulting low-energy effective theory often contains several hundred \cite{Douglas:2006es}. Including all the degrees of freedom of string theory would make our analysis intractable. Instead, we will study 4D  models of $N$ axions $\theta^i$ defined by actions of the form
\be
S=\int d^4 x \, \sqrt{-g}\left({\M^2\over 2}{\cal R} -{1\over 4}F^2+\alpha {{\bs q}_{F\theta}\bs \theta\over 8\pi}F\tilde{F}+{\mathcal L}_{\text{axion}}\right) \,, \label{actionfull1}
\ee
where bold symbols represent vectors or matrices, and matrix multiplication is implicit. The axions are arranged into an $N$-vector $\bs \theta$,  ${\bs q}_{F\theta}$ contains $N$ integer couplings of the axions to a  U(1) gauge field with field strength $F$ (and dual field strength $\tilde{F}$), $\M$ is the reduced four-dimensional Planck mass, and ${\mathcal L}_{\text{axion}}$  is given by
\be\label{lagrthetasec2}
{\mathcal L}_\text{axion}={1\over 2} \partial\bs\theta^\top \bs K \partial\bs\theta - \sum_{I=1}^P \Lambda_I^4  \left[1-\cos\left({\mathbfcal Q}^I\bs\theta\right) \right] -  V_{0} \,.
\ee
Here $\bs K$ is an $N\times N$ metric on the axion field space, ${\mathbfcal Q}$ is a $P \times N$  matrix with integer entries, and the couplings $\Lambda_J^4$ depend on the action of the $J$th instanton like  $\Lambda_J^4\propto e^{-S_J}$. We choose the ``$1 - \cos$" form for convenience so that the non-trivial part of the potential is non-negative, and include  a constant  $V_0$ that represents the minimal possible vacuum energy density.  The cosines could be replaced with another  periodic function without qualitatively affecting our conclusions.  We have ignored the $P$ phases that should in general be included in the arguments of the cosines; these generically can be set to negligibly small values by a shift in the axion fields \cite{bejk2}.

In our recent work \cite{bejk1,bejk2,bejk3} we introduced a  framework that renders the analysis of the  theories \eqref{lagrthetasec2} tractable, even when $N\gtrsim 100$, provided $P \gtrsim N$. When the charges, couplings and field space metric in  \eqref{actionfull1} and \eqref{lagrthetasec2} are chosen randomly, we showed \cite{bejk2} that there are typically ${\cal N}_{\text{vac}} \sim 10^N$ distinct vacua with a smooth distribution of vacuum energies, making an anthropic solution to the CC problem potentially viable in a single instance of a random axion landscape with $N \sim 100$.

In \cite{bejk3} we found that a large fraction of the ${\cal N}_{\text{vac}}$ minima (including those with approximately vanishing CC) are meta-stable and generically have lifetimes longer than the current age of our universe. The neighborhoods surrounding small energy local minima typically contain a region suitable for slow roll inflation.   These inflationary trajectories generate scalar and tensor perturbations with an amplitude and tilt that are close to consistent with observation.

These landscapes can also contain a light direction that is a candidate for ``fuzzy dark matter" \cite{Hu:2000ke}. This occurs when the leading axion potential, which neglects quantum gravity effects, has at least one would-be flat direction in the axion field space (mathematically the condition for this to occur is that the rank of the charge matrix $\mathbfcal{Q}$ describing that potential in \eqref{lagrthetasec2} is less than $N$, the number of axions).  A conjecture in quantum gravity predicts that a small potential will be generated for those directions \cite{Rey:1989mg,ArkaniHamed:2006dz,Rudelius:2014wla,Bachlechner:2015qja,Alonso:2017avz,Hebecker:2018ofv}.  The abundance of the resulting dark matter satisfies $\rho_{\text{DM}}\approx \M^4 \, e^{-\mathcal{O}(1)\times \M/f}$. For a GUT scale decay constant, $f\approx 10^{-2} \M$, this coincides roughly with the observed energy density $\rho_{\text{DM}}\approx \rho_{\text{DE}}$, similar in flavor to the WIMP miracle \cite{Arvanitaki:2009fg, Hui:2016ltb}. Coupling to a U(1) gauge field as in \eqref{actionfull1} does not affect these conclusions.

This analysis suggests that a universe rather like our own can be accommodated in random axion landscapes. What is truly remarkable  is the power of the anthropic requirement of structure formation. Assuming only that the initial state is eternal inflation in a typical meta-stable false vacuum, and requiring that  density perturbations    reach ${\cal O}(1)$, a generic cosmological history  coincides at least roughly with  the entire known expansion history of our universe:  a big bang (here, tunneling from an eternally inflating parent phase), followed by slow roll inflation with sufficiently many efolds to solve the horizon and flatness problems, followed by reheating to a phase of radiation domination, followed by matter domination by a particle consistent in abundance and characteristics with data, followed by a period of vacuum energy domination with a value consistent with data.

All this occurs without tuned small parameters, or any model-building to fit input from observation (apart from the minimal anthropic requirement of structure formation).  As we will see, these cosmologies have primordial density perturbations of order $10^{-4}$-$10^{-5}$ produced during roughly 60 efolds of slow roll inflation.  Moreover, matter-radiation equality occurs at roughly the same energy density as in our universe, and the value of dark energy is parametrically determined by this density to a value similar to that of $\rho_\text{DE}$.  

In some detailed respects  these cosmologies  are  in conflict with observation -- for example we typically find observable tensor modes ($0.07\lesssim r\lesssim 0.14$) as well as detectable isocurvature fluctuations. We could avoid these conflicts with minor modifications of the theory, but for simplicity in this paper we will focus on typical predictions of a simple ``benchmark" model.

\section{The Axidental Universe in a nutshell}\label{nutshell}
The action \eqref{actionfull1} depends on $P$ energy scales $\Lambda_I$, $N(N+1)/2$ real entries in the symmetric and positive definite metric ${\bs K}$, and $N(P+1)$ integer charges contained in ${\mathbfcal Q}$ and  ${\bs q}_{F\theta}$. We take ${\mathbfcal Q}$ and ${\bs q}_{F\theta}$ to contain random independent, identically distributed integers and will denote the root-mean-square of the entries by $\sigma_{\mathcal Q}$ and $\sigma_q$ respectively. For simplicity, in this paper we only consider only kinetic matrices with trivial structure: $ K^i_{~j} = f^2\delta^{i}_{~j}$.

The energy scales $\Lambda_I$ are produced by non-perturbative effects, $\Lambda_I^4\propto e^{-S_I}$. This exponential dependence implies that the $\Lambda_I$ can range over a very wide scale. For non-perturbative corrections deriving from stringy physics, the instanton actions $S_I$ involve integrals over compact cycles of the internal manifold.  If this manifold is compactified at the string scale, one expects  $S_I$ of order one and  $\Lambda_I$ near the string scale as well.  However, under special circumstances stringy effects may not contribute to the potential for one or more directions in the axion field space.  In this case \emph{subleading} non-perturbative terms from quantum gravity become important. The weak gravity conjecture \cite{ArkaniHamed:2006dz} extended to axions places a lower bound on the scale of these subleading terms \cite{Rudelius:2014wla,Cheung:2014vva}. 

In this paper we will make a simple choice:  $P-1$  terms with  equal energy scales $\Lambda_J=\Lambda$ (for $J \leq P-1$) that lift all but one of the axion directions.\footnote{A more general discussion can be found in our previous works, cf. Section 4 of \cite{bejk3} or Appendix 6 of \cite{bejk2}.}
  With some foresight, we label the $P$th energy scale $\Lambda_P=\Lambda_{\text{DM}} \ll \Lambda$.\footnote{We are free to re-order the rows of ${\mathbfcal Q}$, so we can choose $\Lambda_P=\Lambda_{\text{DM}}$ without loss of generality.}   Roughly saturating the bound from the weak gravity conjecture corresponds to $\Lambda_{\text{DM}} \sim \M \, e^{-\mathcal{O}(1) \times \M/f}$.  This produces a very small mass for the would-be massless axion, making it a natural fuzzy dark matter candidate.

\subsection{Benchmark model}
Throughout this paper we will restrict attention to a simple benchmark model with the following parameters:
\bea \label{benchmark}
N & = & P-1= 500 \,, \notag \\ 
f &=& \Lambda_{I \neq P} =  \Lambda= (-V_0)^{1/4} = 1.0 \times 10^{-2} \M \,, \notag \\
\Lambda_{\text{DM}}^4 &= & \Lambda_P^4= \M^4 \, e^{-{\cal S}\M / f}, \, {\cal S}= 2.3  \,, \\
\sigma^2_{\mathcal Q}  &=&  \sigma^2_q  = 2/3, \,\,\, \alpha  =  0.1 \,. \notag
\eea   
We choose all the integer charges (the elements of the matrix ${\mathbfcal Q}$ and the vector ${\bs q}_{F\theta}$) to be $\pm 1$ or $0$ with equal probability, so that their variances are $ \sigma^2_{\mathcal Q}  =  \sigma^2_q  = 2/3$. 
The infrared energy scale  $\Lambda_{\text{DM}}$ is discussed above and  in \S\ref{radandmatter}, and    $\cal S$ denotes an $\mathcal{O}(1)$ factor that is not precisely known.

We choose  $P -N=1$ because our results are simpler to state in that case, but nothing substantial changes in the range $2N \gg P > N+1$ \cite{bejk2}. We choose $N=500$ because it is a round number characteristic of the numbers of axions in string compactifications \cite{Douglas:2006es}, and because with this choice the number of vacua turns out to be ${\cal N}_{\text{vac}} \sim 10^{500}$ (a favorite number of ours).
We choose the dimensionful parameters at $\sim 10^{-2} \M$ because it is close to the GUT scale, and to the string scale in traditional models of string phenomenology, so it is a reasonable guess for the scale where a large number of axion fields might appear.  Setting $\Lambda$, $f$, and $-V_0^{1/4}$ to \emph{precisely} $1.0 \times 10^{-2} \M$, and choosing the entries of ${\mathbfcal Q}$ and ${\bs q}_{F\theta}$ to be $\pm 1$ or $0$ with equal probability, are simple choices meant to illustrate the lack of sensitivity of our general conclusions to the precise values of these parameters. The dimensionless coupling $\alpha$ must be somewhat less than one (or else the interaction with photons will strongly affect inflation), but otherwise the results are not very sensitive to it, so we set it to $0.1$ (a smaller value would simply lower the reheating temperature).  Our results \emph{are} very sensitive to the combination ${\cal S} \M/f$ because $\Lambda_{\text{DM}}$, defined in Eq. \eqref{benchmark}, depends exponentially on it.  The value of $\cal S$ is expected to be $\mathcal{O}(1)$; as we will see the choice  ${\cal S}= 2.3$ produces universes very similar to ours.

\section{Initial conditions and tunneling}
The initial conditions of our universe -- if they exist -- are unknown to us. One compelling approach to this problem has been inspired by the hope that the detailed initial conditions may be irrelevant to observations. We will see that this hope is plausibly realized in the theory (\ref{actionfull1}).

The potentials of multi-axion theories typically have a vast number of discrete, meta-stable minima at field space locations $\bs \theta_{\text{vac},n}$, where the index $n=1,\dots,{\cal N}_\text{vac}$ labels the minimum. The history of any part of the universe will spend the vast majority of its  time close to these meta-stable minima, with relatively brief transient periods in which the axions, gravity, and gauge-fields are dynamical.

Let us briefly review the vacuum distribution in the theory (\ref{actionfull1}), assuming the benchmark parameters. The number of discrete vacua scales super-exponentially with the number of axions.  When the number of relevant non-perturbative terms in the axion potential exceeds the number of axions by one, $P=N+1$, we can analytically approximate the vacuum locations and energy densities. We  have
\be\label{minimapnplus1}
V(\bs \theta_{\text{vac},n}) \approx {2\pi^2}\Lambda^4 {{ \, n^2\over \det(\mathbfcal Q^\top \mathbfcal Q)}} +V_0 \,.
\ee
 The determinant of the $N \times N$ matrix $\mathbfcal Q^\top \mathbfcal Q$ rapidly becomes extremely large with $N$, and so from \eqref{minimapnplus1} the vacuum energies become very closely spaced.  The number of vacua scales as the square root of the determinant:
\be
{\cal N}_{\text{vac}}\sim  {\sqrt{\det (\mathbfcal Q^\top \mathbfcal Q)}} \sim \sigma_{\mathcal Q}^P \sqrt{P!} \approx 10^{524}
\ee
where we have dropped various subleading factors \cite{bejk2} and used \eqref{benchmark}.\footnote{It is interesting to note that the large value of ${\cal N}_{\text{vac}}$ can be thought of as arising from a single integer (in general, $P-N$ integers) with a huge range, $1 < n <  {\sqrt{\det (\mathbfcal Q^\top \mathbfcal Q)}}$ \cite{bejk2}, rather than from the combinatorics of multiple integer choices with relatively small ranges as in \cite{Bousso:2000xa}.}

Each minimum has on the order of $3^P\approx 10^{238}$ neighboring minima with vacuum energy densities distributed approximately uniformly between $V_0$ and $V_0+0.14\times P\Lambda^4$ \cite{bejk2}. Since $V_0 \sim -\Lambda^4$ is negative, every minimum will have a vast number of neighbors with vacuum energy densities $ | \rho_\text{vacuum}| <  10^{-120}\M^4$.

In a theory with multiple minima there exists a natural mechanism that populates minima and potentially sets up inflationary initial conditions -- Coleman-de Luccia (CdL) tunneling from a ``parent" false vacuum \cite{Coleman:1980aw}. Indeed, this was the first version of inflation proposed by Guth \cite{Guth:1980zm}, although in that model no inflation took place after tunneling.  Instead,  tunneling in our benchmark landscape is sometimes followed by a substantial amount of slow roll inflation.

The initial conditions inside a bubble after tunneling are those of a strongly negatively curved, approximately homogeneous FLRW cosmology, with metric
\be\label{flrwmetric}
\mathrm{d}s^2 = \mathrm{d} t^2 - a(t)^2 \left( \mathrm{d}r^2+\sinh^2(r) \mathrm{d} \Omega_2^2\right) \,.
\ee
At $t=0$ the axions take some initial values $\bs \Theta_{t=0}$ (where $\bs \Theta$ denotes a canonically normalized choice of axions), and the universe is  dominated by negative curvature due to the initial condition $a(t) \sim t$ as $t \rightarrow 0$.\footnote{It is interesting to note that $t=0$ is a coordinate singularity -- the metric is  smooth there.} The Friedmann equations are
\be
H^2={\rho\over 3\M^2}-{\kappa\over a^2} ~~~\text{and}~~~{\ddot{a}\over a}=-{\rho+3p\over 6\M^2} \,,
\ee
where $\kappa=-1$ because the universe is open, $H=\dot{a}/a$ is the Hubble parameter, and $\rho$ and $p$ denote the energy density and pressure, respectively. We expect the vast majority of tunneling events to result in non-inflationary universes.  However,  if by chance the initial condition after tunneling is on a part of the potential that can support slow roll inflation with $\rho_{\text{inf}} \approx -p_{\text{inf}}$, the potential energy begins to dominate after a time
\be
t_{\kappa\rightarrow\text{inf}}\approx \sqrt{3\M^2\over \rho_{\text{inf}}}=H^{-1}_{\text{inf}} \,.
\ee
Since the scale of the potential is set by $\Lambda$, a  rough estimate for  the Hubble parameter during inflation, $H_\text{inf}$, is
\be\label{einf}
H^2_{\text{inf}}\sim \Lambda^4/\M^2 \,.
\ee
(See \cite{bejk3} for a more detailed discussion of inflation in these landscapes.)

To summarize, while the vast majority of cosmological histories after tunneling have very large CC and no prolonged period of slow roll inflation, there do exist cosmological histories that could support high-scale inflation and a small CC. We will see below that only these latter histories lead to structure formation.

\section{Inflation and reheating}\label{infreh}
Whether an extended period of inflation (see e.g. \cite{Baumann:2009ds}) occurs depends on whether the slow roll parameters are small, $\epsilon,\,\eta\ll1$, where
\be
\epsilon \equiv {\M^2\over 2} \left(V'\over V\right)^2\,,~~~~~\eta \equiv { \M^2}{V''\over V} \,,
\ee
and primes denote derivatives with respect to the canonically normalized inflaton.  The slow roll parameters will not typically be small for  a random initial value of the axion fields, but the fields rapidly fall down the steep directions of the potential, leaving them displaced from the nearest minimum along the most gently sloped directions.
 Once that happens, the number of efolds of expansion during the inflationary phase is related to the inflaton field displacement $\Delta \bs \Theta$ by 
\be
N_e \approx {1 \over 2} \left({\Delta\bs \Theta\over\M}\right)^2 \,,\label{approxefolds}
\ee
where the $1/2$ is valid for a roughly linear potential. (The inflationary potentials in our benchmark model tend to interpolate between linear and quadratic.)

The inflaton displacement $\Delta \bs \Theta$ cannot be much larger than the largest scale over which the axion potential is roughly linear or quadratic. As discussed in \cite{bejk2}, the features of multi-axion potentials are highly anisotropic, giving rise to directions in the field space that are much smoother than expected from naive dimensional analysis. This qualitative phenomenon is known as (kinetic or lattice) axion alignment \cite{Kim:2004rp, Bachlechner:2014hsa,Bachlechner:2014gfa,bejk2}. Some amount of alignment is generic in  multi-axion theories, giving rise to a relatively flat direction within the potential. For the  ensemble of theories considered here, for a typical minimum the field range in this flattest direction is given in Eq. (4.30) of \cite{bejk2},
\be\label{inflatondisp}
\Delta{\bs \Theta} \sim  \frac{\pi \, f }{  \ell(P)  \sigma_\mathcal{Q} \left( 1 - \sqrt{N/P} \right) }\approx 12\M\,,
\ee
 where we used our benchmark parameters \eqref{benchmark}: $N = 500, P=501$ and $f= 1.0 \times 10^{-2} \M$.  Comparing with (\ref{approxefolds}), the number of efolds near a typical minimum is\footnote{Certain minima for any given theory \eqref{lagrthetasec2}, and generic minima for certain choices of the data  $\mathbfcal{Q}$ and $\bs K$, will allow for substantially more efolds than this.  Also, more efolds of inflation may occur at approximate saddle points. However, the initial conditions following Coleman-de Luccia decay from tunneling the nearest minima do not typically access this full range  \cite{bejk2, bejk3}.}
\be\label{efolds}
N_e\lesssim 72 \,.
\ee

Inflation proceeds along some direction $\hat{\bs \theta}_{\text{inf}}$ in axion field space. The inflaton couples to the gauge field $F$ via the coupling\footnote{We can estimate the typical effective scale $f_{\text{inf}F}$ at which the inflaton couples to the gauge field from (\ref{actionfull1}).  Assuming inflation occurs along the lightest direction $\hat{\bs \psi}$, which typically coincides with the eigenvector of $\mathbfcal{Q}^\top \mathbfcal{Q}$ with smallest eigenvalue, we have $\bs \theta_\text{inf} \approx (\Theta_\text{inf} / f) \, \hat{\bs \psi}$ (recall $\bs \Theta$ stands for a canonically normalized axion). Since $\hat{\bs \psi}$ is \textit{delocalized} to a high degree \cite{Bachlechner:2014gfa,bejk2}, we have $|{\bs q}_{F\theta} \hat{\bs \psi}| \approx \sigma_q $ and so ${\bs q}_{F\theta} \bs \theta \approx (\sigma_q/f ) \Theta_\text{inf} \equiv \Theta_\text{inf} / f_{\text{inf}F}$.}
\be
{\alpha\over8\pi f_{\text{inf}F}}\Theta_\text{inf} F\tilde F \,,
\ee
where $f_{\text{inf}F}\approx f/\sigma_q = 10^{-2} \M$ and $\alpha= 0.1$. When $\Theta_\text{inf}$ is evolving slowly or not at all, this coupling is topological and does not affect the dynamics. However when inflation ends  the inflaton begins rolling and/or oscillating rapidly, at which point this coupling can efficiently transfer energy from the inflaton to the gauge field and reheat the universe \cite{Adshead:2015pva}.  The dimensionless parameter that controls the energy transfer is
\be
\xi={\alpha\M\over 2\pi f_{\text{inf}F}}\sqrt{\epsilon\over 2} \approx 0.9 \sqrt{\epsilon} \,,
\ee
where $\epsilon$ is the slow roll parameter and the approximation is valid with our  benchmark parameters.  At the end of inflation, energy is transferred  to (one helicity state of) the gauge field.

We can obtain a lower bound on the reheating temperature by comparing the expansion rate of the universe $H$ after inflation to the perturbative decay rate of axions into photons  \cite{Adshead:2015pva}
\be
\Gamma_{\Theta_\text{inf}\rightarrow F}={\alpha^2\over 512\pi^3 f_{\text{inf}F}^2} m_{\Theta_\text{inf}}^3 \,.
\ee
The inflaton mass is roughly given by (see e.g. (3.24) of \cite{bejk2})
\be
m_{\Theta_\text{inf}}^2\approx {\Lambda^4 \sigma_{\mathcal{Q}}^2 \over N f^2}\approx  \mathcal{O}(10^{-7})\M^2 \,,
\ee
where we substituted the parameters of our benchmark model in the last equality. Solving $\Gamma_{\Theta_\text{inf}\rightarrow F}\le H_{\text{reh}}=\sqrt{T_{\text{reh}}^4/3\M^2}$, and given our benchmark parameters, we have a lower bound on the reheating temperature, $T_{\text{reh}}\gtrsim 10^{-6}\M$. 
Hence, the reheating temperature is somewhere within the interval
\be \label{Tre}
10^{-6}\M\lesssim T_\text{reh}\lesssim 10^{-2}\M \,,
\ee
where the upper bound is the energy scale of inflation \eqref{einf}.  A detailed study of the non-perturbative reheating dynamics at the end of inflation would be needed to give a more precise answer. 

As a conservative choice (in that structure formation will require more inflation) we will assume a  reheating temperature near the upper limit.  Choosing the lower bound would not affect our qualitative findings.  Lowering the reheating temperature by  $10^4$  reduces the number of efolds required to solve the horizon problem  by roughly $\log 10^4 \approx 9$.

As is well known, slow roll inflation generates an almost scale-invariant and Gaussian spectrum of perturbations. In the comoving gauge ($\delta \bs \theta_\text{inf}= \bs 0$), the metric (\ref{flrwmetric}) is perturbed by
\be
\delta g_{ij}=a^2(1-2\zeta)\delta_{ij}+a^2 h_{ij} \,,
\ee
where $h_{ij}$ parametrizes the transverse and traceless tensor perturbations, while $\zeta$ denotes the scalar perturbations.  It is convenient to expand $\zeta$ in  modes of  comoving wave-numbers $k=a/\Lambda$, where $\Lambda$ is the physical wavelength.  The modes $\zeta_k$ are conserved outside the horizon:\footnote{Assuming purely adiabatic perturbations.}  $\dot{\zeta}_k=0$ for $k\ll a H$. After quantizing, the power spectrum for $\zeta$ is ${2\pi^2}\Delta_\zeta^2(k)/k^3$, where
\be\label{pwrspectrum}
\Delta_\zeta^2(k)={1\over 8\pi^2} {H_\text{inf}^2\over \epsilon\M^2}\bigg|_{k=a H} \,.
\ee

We can roughly estimate the power spectrum as follows. Assuming an approximately linear potential, $\epsilon$ is  related to the number of efolds $N_e$ before the end of inflation by $\epsilon\approx {1/ 4N_e}$. Using the scale of inflation (\ref{einf}), together with (\ref{approxefolds}) and (\ref{inflatondisp}) we  have
\be \label{perts}
\Delta_\zeta(k)\approx{\sqrt{N_e}  \over \sqrt{6}\pi}{ H_\text{inf} \over \M}\approx 1\times 10^{-4} \,.
\ee
Our crude benchmark model comes surprisingly close to the observed values for the primordial curvature fluctuations of $\Delta_\zeta|_{\text{observed}}\approx 5\times 10^{-5}$.

To conclude this section, we have now established that there typically exist inflationary trajectories that yield tens of efolds of expansion and terminate in reheating the Abelian gauge field at the end of inflation to a high energy density. This high-temperature (thermalized \cite{Ferreira:2017lnd}) photon gas has small density perturbations that are roughly consistent with those observed in our universe. 

\section{Radiation and matter domination}\label{radandmatter}
Inflation ends and the universe reheats at time $t_\text{reh}\approx (N_e+1) H_{\text{inf}}^{-1}$.  For simplicity we assume instant reheating to a  temperature  $T_{\text{reh}} \approx 10^{-2} \M$ at the top of the range \eqref{Tre}, but as mentioned above our qualitative conclusions are not sensitive to $T_{\text{reh}}$. 

The primordial density perturbations generated during inflation  \eqref{perts} are very small, and  grow only logarithmically with the scale factor during radiation domination, and not at all during curvature or vacuum energy domination. Therefore,  cosmologies in which structure forms must go through a period of matter domination (when perturbations grow roughly linearly with the scale factor).   For matter to dominate at any time, matter-radiation equality must occur before either curvature or vacuum energy dominate.\footnote{This assumes the vacuum energy is positive.  If \emph{negative} vacuum energy dominates at time $t_\text{vac}$, the universe will collapse to a big crunch at a time of order $2 t_\text{vac}$.  While it is conceivable that non-linear structures could form during this collapsing phase, this provides at most a factor of a few in our bounds and so for simplicity of the presentation we will ignore this possibility.} As we will see, in our benchmark model the energy density in matter is tiny.  It is this that ultimately gives rise to the tiny value of dark energy (in cosmologies that form structure).

In order to estimate the matter density, we parametrize the mass of the lightest axion by 
\be
m_{\text{light}}\approx {\Lambda_{\text{DM}}^2\over f}\approx {\M^2\over f} e^{-{\cal S} \M/2f} \,,
\ee
where ${\cal S}=\mathcal{O}(1)$ is a parameter we set to $2.3$ in our benchmark model. As mentioned in \secref{nutshell}, such tiny scales  can be generated by non-perturbative gravitational effects \cite{Rey:1989mg,ArkaniHamed:2006dz,Rudelius:2014wla,Bachlechner:2015qja,Alonso:2017avz,Hebecker:2018ofv}. In fact, the  scale $\Lambda_{\text{DM}}^4\approx \M^4 \, e^{-\M/f}$ is roughly what one expects for a would-be massless axion if the weak gravity conjecture extended to axions is saturated.  

With our benchmark parameters \eqref{benchmark}, this  gives a dark matter mass of $m_{\text{light}}\approx 3\times 10^{-21}~\text{eV}\approx 10^{-48}\M$, a typical value for fuzzy dark matter \cite{Hui:2016ltb}. The Hubble scale at reheating is far above this mass (even if the lower bound on reheating temperature is chosen).  Therefore the light axion remains frozen by friction until the Hubble scale drops below its mass, at which point it begins to oscillate.  This occurs  when
\be
m_{\text{light}}\approx H(t_\text{osc}) \approx H_\text{inf} \left({a(t_\text{osc})\over a(t_\text{reh})}\right)^{-2} \,,
\ee
where we assumed radiation domination from reheating to $t_\text{osc}$.  Using $a(t_\text{reh})\approx H_{\text{inf}}^{-1}e^{N_e}$, the scale factor at $t_\text{osc}$ is
\be
a(t_\text{osc})\approx \left({H_{\text{inf}}\over m_{\text{light}}}\right)^{1/2} H_{\text{inf}}^{-1}e^{N_e}\approx 5 \times 10^{21} a(t_{\text{reh}})\approx10^{26}e^{N_e}\M^{-1} \,.
\ee
The energy density in the light axion before it begins to oscillate is of order 
$$
\Lambda_{\text{DM}}^4\approx m_\text{light}^2 f^2\approx  10^{-100}\M^4 \,.
$$
Because the potential for the light axion is approximately quadratic, once it begins to oscillate it behaves like cold dark matter with equation of state $w=0$\footnote{The negative quartic corrections due to the cosine can actually have a non-negligible effect and may ameliorate  Lyman-$\alpha$ constraints \cite{Desjacques:2017fmf}.}
\be\label{energydensityaxion}
\rho_\text{light}(t)=\Lambda_{\text{DM}}^4\left({a(t)\over a(t_\text{osc})}\right)^{-{3}}~~~\text{for}~~t>t_\text{osc} \,.
\ee
The universe remains radiation dominated until matter-radiation equality $\rho_\text{light}(t_{\text{eq}})\approx \rho_{\text{rad}}(t_{\text{eq}})$:
\be
a(t_{\text{eq}})\approx {T_{\text{reh}}^4\over \Lambda_\text{DM}^4}{a(t_\text{reh})^4\over a(t_\text{osc})^3}\approx 4\times 10^{30}e^{N_e}\M^{-1} \,.
\ee

\begin{figure}
  \centering
  \includegraphics[width=1\textwidth]{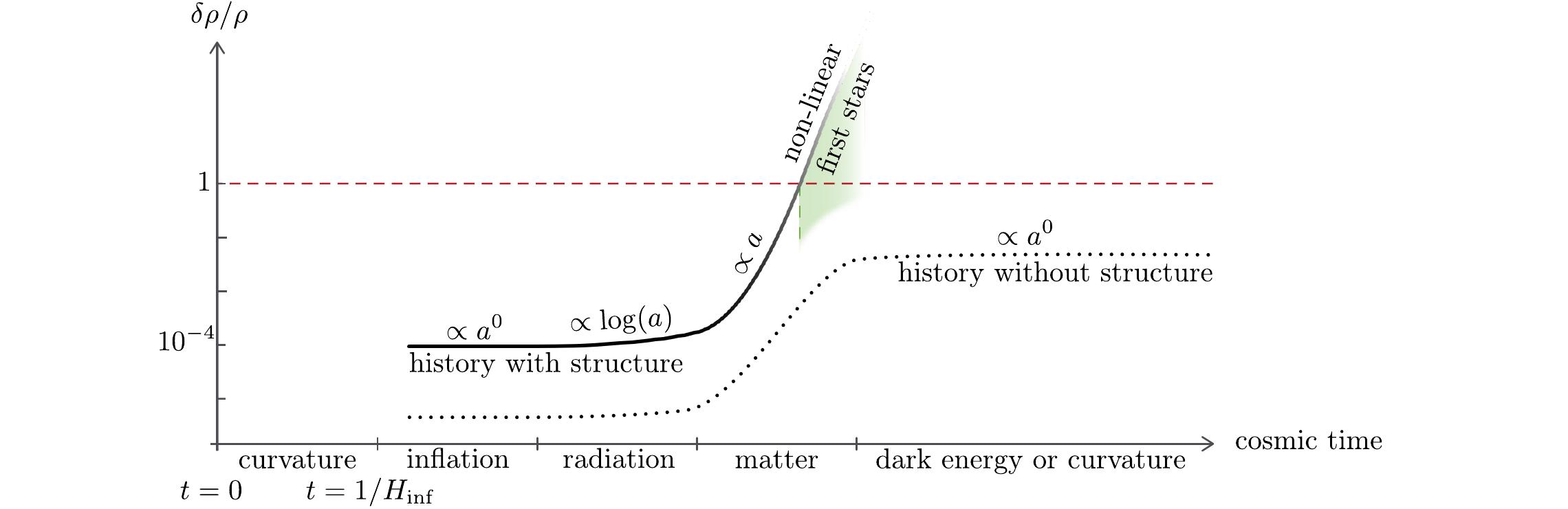}
  \caption{\small Schematic evolution of density perturbations over time for a cosmology that forms non-linear structure (solid line) and a cosmology that remains structureless (dotted line).}\label{expansionhistory}
\end{figure}

To summarize, after the end of inflation and reheating into radiation, a cold dark matter component with a tiny energy density is formed when the light axion begins to oscillate. Since radiation dilutes faster than matter, the universe eventually becomes   matter dominated. Density perturbations grow efficiently only during matter domination, and so the universe has to remain matter dominated until structure forms, and only then can either curvature or vacuum energy can take over.  This requirement forces dark energy to take a tiny value, and requires a sufficient number of efolds of inflation to make the curvature tiny as well.  We schematically illustrate this in Fig.~\ref{expansionhistory}.

\subsection{Matter density at structure formation} \label{matterdensitysec}
Let us now discuss the growth of perturbations during matter domination to find the scale factor at which structure forms. By ``structure  forms" we mean that the perturbation reaches $\mathcal{O}(1)$ on some length scale. If either curvature or vacuum energy dominate prior to matter-radiation equality, perturbations do not grow and no structure can form, and even if there is a period of matter domination it must be sufficiently long.  Lastly, the mass of the dark matter particle must be large enough that it begins to oscillate (and cluster) prior to these times.  

The light axion field is frozen before $t_\text{osc}$, but begins to behave like  a pressureless fluid thereafter. The curvature perturbations, which have variance $\Delta_\zeta^2$, are scale invariant and imprinted onto the matter density perturbation $\delta_m=\delta\rho_m/\rho_m$ at $t_\text{osc}$ with variance (see e.g. \cite{Baumann:2009ds})
\be
\langle\delta_m^2(k,a_{\text{osc}}) \rangle\approx \Delta_m^2|_{t_\text{osc}}=2\pi^2 {4\over 25}\Delta_\zeta^2|_{t_\text{osc}} \,.
\ee
As mentioned above, super-horizon matter perturbations with wave-numbers $k<a H$ are frozen, while sub-horizon modes grow logarithmically during radiation domination, $\delta\propto \log(a/a(t_\text{eq}))$, and grow linearly, $\delta\propto a$, during matter domination. We are interested in the time of structure formation, i.e. the time at which the first sub-horizon perturbation becomes large $\delta_m>1$. 

Let us consider a mode with wavenumber $k$ that is inside the horizon when the light axion begins to behave like cold dark matter, $k>a(t_{\text{osc}}) H(t_{\text{osc}})$. The scale of this perturbation grows logarithmically up to $a(t_\text{eq})$, and linearly thereafter, so during matter domination we roughly have
\be
\delta_m(k>a(t_{\text{osc}}) H(t_{\text{osc}}),a>a(t_{\text{eq}}))={a\over a(t_\text{eq})}\log\left({a(t_{\text{eq}})\over a(t_\text{osc})}\right)\Delta_m|_{t_\text{osc}} \,.
\ee
At equality, the matter perturbations roughly are of order $\delta_m|_{t_\text{eq}}\approx 2\times 10^{-3}$. Large matter perturbations therefore occur at a time
\be\label{adeltag1}
a|_{\delta_m>1}={a(t_\text{eq})\over \log\left({a(t_{\text{eq}})\over a(t_\text{osc})}\right)} \Delta_m^{-1}|_{t_\text{osc}}\approx e^{76.6+N_e}\M^{-1}H \,.
\ee
The matter density at structure formation is given by\footnote{This is roughly an order of magnitude smaller than the dark matter density in today's universe, which in turn is smaller than the density when the first structures formed.  Hence our simple benchmark model forms structure  at a smaller density than  our universe.  However, as mentioned above it would require only a relatively small adjustment of the parameters  to correct this.}
\be\label{rhomstructure}
\rho_m(a|_{\delta_m\approx1})\approx \Lambda_{\text{DM}}^4 \left({a(t_\text{eq})\over a(t_\text{osc}) }\right)^{-4}\left({a|_{\delta_m\approx1}\over a(t_\text{eq}) }\right)^{-3}\approx 3\times  10^{-122}\M^4 \,.
\ee

\subsection{Basic constraints from structure formation}
With (\ref{rhomstructure}) we have found the dark matter density at the time when non-linear structures can begin to form, but only if the curvature and vacuum energy are negligible when the scale factor reaches $a|_{\delta_m\approx1}$.  Hence the requirement that structure forms translates to a very simple constraint on the duration of inflation and the CC,
\be
| \rho_\text{vacuum} | \lesssim \rho_m(a|_{\delta_m\approx1})\approx 3\times 10^{-122}\M^4\,,~~\Omega_k|_{\delta_m\approx 1}={1\over (aH)^2}\bigg|_{\delta_m\approx 1}\lesssim 1 \,,
\ee
where with (\ref{adeltag1}) and (\ref{rhomstructure}) the second inequality implies a lower bound 
$$N_e\gtrsim 64$$ 
on the amount of efolds of inflation. This  agrees well with a more sophisticated analysis following \cite{Weinberg:1987dv,Freivogel:2005vv}, that we present in Appendix \ref{app1}.\footnote{This more detailed analysis gives the slightly stronger bounds $\rho_\text{vacuum}\lesssim 2\times 10^{-122}\M^4$ and $N_e\gtrsim 63.5$.}

\section{The observable universe}
Life, and computation, becomes complicated after non-linear structures form. In order to still get a rough idea of what a typical observer in such a universe might observe, we now assume an $\mathcal{O}(1)$ amount of expansion occurs between the onset of structure formation and the scale factor of observations, $a_\text{``today''}\approx \mathcal{O}(1) a|_{\delta_m\approx1}$.

The vacuum energy density is constrained to be below $10^{-122}\M^4$ in absolute value. Our theory has a vast number of vacua that satisfy this constraint, and the vacua are distributed uniformly to a very good approximation within the constraint. This gives the expected value of the vacuum energy density
\be
| \rho_\text{vacuum} |= \mathcal{O}(10^{-122})\M^4 \,.
\ee
At the time when observations are made, dark energy (or curvature) will be dominant and the matter abundance will have decreased somewhat to
\be
\Omega_\text{m}= \left({1\over \mathcal{O}(1)}\right)^3 \,.
\ee

Similarly, we can give a rough  prediction for the curvature abundance $\Omega_k$.  Let us assume that the number of efolds among cosmological histories is distributed  uniformly between the minimal number required for structure formation in (\ref{minefolds}) and the upper bound (\ref{efolds}).  Then the prediction for the number of efolds is
\be
N_e=68\pm 3 \,.
\ee 
The  spatial curvature depends on the time when the observation is made.  If we choose the time when $\Omega_\text{m} = 0.3$, our estimates give a range
\be
-6.6 < \log_{10}(|\Omega_k|) < 1.4 \,.
\ee
The current observational bound is $|\Omega_k|\le 0.005$, and the in-principle best possible bound  is $|\Omega_k| \simleq 10^{-4}$ \cite{Kleban:2012ph}. 

The relative scale factor between matter-radiation equality and today is
\be
{a_{\text{``today''}}\over a(t_\text{eq})}={a|_{\delta_m\approx1}\over a(t_\text{eq})}\times\mathcal{O}(1)\approx 500 \times \mathcal{O}(1) \,,
\ee
similar to the observed value. 

\section{Conclusions}
In this work we studied the cosmological histories that arise in a  theory of several hundred axions  minimally coupled to gravity and to an Abelian gauge field. The vast majority of cosmological histories contain no structure; they are empty universes inconsistent with the most basic of observations. The few cosmological histories that do contain structure typically exhibit an expansion history very similar to our own: an extended period of cosmic inflation followed by reheating, radiation domination, matter domination, and ultimately dark energy domination with an exponentially small cosmological constant, $\rho_{\text{DE}} \approx \M^4 \, e^{-\mathcal{O}(100)}$. The underlying reason is simple. Perturbations that form structure can grow only during an era of matter domination, the energy density in matter is very small, and therefore the vacuum energy and curvature must be small enough to allow for  a matter-dominated period.

More specifically, setting all dimensionful parameters to the GUT scale, choosing order one random charges for $N\approx 500$ axions, and assuming one lightest possible direction in field space, we find  inflation that lasts around $60$ efolds, fuzzy dark matter with a mass of $m_\text{DM}\approx 10^{-21}\text{eV}$, and matter-radiation equality at a redshift factor of $z_\text{eq}\approx \mathcal{O}(1)\times 500$. At the time when observations are made, the dark matter abundance is roughly $1/\mathcal{O}(1)^3$, and there is a small (open universe) spatial curvature not far below the current observational bounds.

To summarize:  without any tuning and with parameters that seem generic given our current state of knowledge of fundamental physics, our results show that the \emph{typical} cosmology in which non-linear structures form looks very similar to the universe around us.  This may explain many of the apparently tuned features of our universe.

\section*{Acknowledgements}
The work of MK is supported by the NSF through grants PHY-1214302 and PHY-1820814. The work of KE is supported in part by the INFN.  The work of TB is supported in part by the DOE under grants no. DE-SC0011941 and DE-SC0009919 and by the Simons Foundation SFARI 560536. OJ is supported by a James Arthur Graduate Fellowship.

\begin{appendix}
\section{Structure formation constraints} \label{app1}
We will want to compare the upper bound on the vacuum energy density to the smallest spacing in vacuum energies in our theory, which is given by
\be
\Delta V_\text{vacua}\approx {\Lambda^4\over {\cal N}_\text{vac}}\approx 10^{-532}\M^4 \,. 
\ee
Therefore, there exist a vast number of vacua that satisfy the bound on the CC. In imposing the constraints from curvature and vacuum energy we follow the treatment of Section 3 in \cite{Freivogel:2005vv}.

The evolution of matter perturbations $\delta_m$ follows the Friedmann equation
\be\label{perturbationfriedman}
\dot{a}^2={a^2\over 3\M^2}\left(\rho_m+\delta\rho_m+\rho_0\right)-\kappa-\delta \kappa  \,,
\ee
where $\rho_0$ is the vacuum energy density and $\delta \kappa$ is the curvature perturbation due to $\delta \rho_m$ (for simplicity we are restricting to $t > t_\text{eq}$ and neglecting the contribution from radiation). The matter densities satisfy the conservation equation $a^3(\rho_m+\delta\rho_m)={\cal M}$, for some constant ${\cal M}$. An approximate solution to \eqref{perturbationfriedman} is
\be \label{aaprox}
a(t)\approx \left({3{\cal M} t^2\over4\M^2}\right)^{1\over3}-{1\over 10}\left({9\M^2t^4\over 2{\cal M}}\right)^{1\over3}(\kappa+\delta\kappa) + \mathcal{O}(t^2) \,.
\ee
Evaluating the corresponding conserved matter density yields the matter perturbation
\be
\delta\rho_m={6\over 5}\left({6\M^{10}\over {\cal M}^2 t^4}\right)^{1\over 3} + \mathcal{O}(t^{-2/3}) \,.
\ee
Because the matter density dilutes as $\rho_m\propto a^{-3}$,  the combination $\delta\rho_m^3/\rho_m^2$ is constant at early times near $t_\text{eq}$ when the first term \eqref{aaprox} dominates.  

For the overdensity to lead to gravitational collapse, the scale factor must reach a maximum $\dot{a}=0$. Minimizing the right-hand side of (\ref{perturbationfriedman}), we can write the condition $\dot{a}=0$ as a bound on curvature and background energy density:
\be\label{mainconstraint}
\rho_0^{1/3}-{2^{2/3}\M^2\over \rho_m^{2/3}a^2}\kappa\le\left({500\over 729}{\delta\rho_m^3\over \rho_m^2}\right)^{1\over 3} \,.
\ee
For negative spatial curvature ($\kappa=-1$) this is a lower bound on the scale factor $a$, and thus on the number of efolds of inflation $a\propto e^{N_e}$, as well as a bound on the vacuum energy density $\rho_0$. We now discuss each of the constraints in turn.

In order to obtain a constraint on the number of efolds, we neglect $\rho_0$ and set $\kappa=-1$ in (\ref{mainconstraint}). The scale factor at matter-radiation equality is conveniently expressed as 
\be
a(t_\text{eq})={e^{N_e}\over \sqrt{H_\text{inf} H(t_\text{eq})}} \,.
\ee
This immediately yields a lower bound on the duration of inflation,
\be\label{minefolds}
N_e\ge{1\over 2}\log \left({3H_\text{inf}\over 5H(t_\text{eq})\delta_m(t_\text{eq})}\right)\approx 63.5 \,,
\ee
where we substituted the benchmark parameters and used $\delta_m(t_\text{eq})\approx 2\times 10^{-3}$, a value we derived in \S\ref{matterdensitysec}.

If we neglect curvature we immediately obtain Weinberg's bound on the vacuum energy density \cite{Weinberg:1987dv}
\be\label{weinbergbound}
\rho_0\le{500\over 729} \rho_m\delta^3\bigg|_{t_\text{eq}}\approx 2\times 10^{-122}\M^4 \,,
\ee
where again we substituted the values for our specific example in the last approximation.

\end{appendix}

\bibliographystyle{klebphys2}
\bibliography{tunnelingrefs}
\end{document}